\documentclass[12pt,reqno,a4paper]{amsart}
\usepackage{extsizes}
\usepackage{amssymb,mathabx}
\usepackage[numbers]{natbib}
\usepackage[dvipsnames]{xcolor}
\usepackage{graphicx}
\usepackage{braket}
\usepackage{soul}
\usepackage[pdftex,plainpages=false,colorlinks,hyperindex,bookmarksopen,linkcolor=red,citecolor=blue,urlcolor=blue]{hyperref}
\bibpunct{[}{]}{;}{n}{,}{,}
\usepackage[hmargin=3cm,vmargin={3.5cm,4cm}]{geometry}

\usepackage{tikz,xcolor,hyperref}

\definecolor{lime}{HTML}{A6CE39}
\DeclareRobustCommand{\orcidicon}{%
	\begin{tikzpicture}
	\draw[lime, fill=lime] (0,0) 
	circle [radius=0.16] 
	node[white] {{\fontfamily{qag}\selectfont \tiny ID}};
	\draw[white, fill=white] (-0.0625,0.095) 
	circle [radius=0.007];
	\end{tikzpicture}
	\hspace{-2mm}
}

\foreach \x in {A, ..., Z}{%
	\expandafter\xdef\csname orcid\x\endcsname{\noexpand\href{https://orcid.org/\csname orcidauthor\x\endcsname}{\noexpand\orcidicon}}
}

\usepackage{multicol}
\numberwithin{equation}{section}

\allowdisplaybreaks

\begin{document}

\title[Hyperbolic-type decay]{On the application of Mittag-Leffler functions to hyperbolic-type decay of luminescence.}

\author{Ambra Lattanzi$^{1,2}$ \orcidA{}}\thanks{ambra.lattanzi@gmail.com \hspace{0.5cm} ambra.lattanzi@ifj.edu.pl (A. Lattanzi)} 
\address{${}^1$H. Niewodniczański Institute of Nuclear Physics Polish Academy of Science IFJ-PAN, Kraków, Poland}

\address{${}^2$ENEA Research Center of Frascati, Frascati, Rome, Italy}

\author{Giampietro Casasanta$^3$ \orcidB{}}\thanks{g.casasanta@isac.cnr.it (G. Casasanta)}
\address{${}^3$Institute of Atmospheric Sciences and Climate, ISAC/CNR.}

\author{Roberto Garra$^4$ \orcidC{}}\thanks{roberto.garra@uniroma1.it  (R. Garra)}
\address{${}^4$Department of Statistical Sciences, Sapienza University of Rome.}


\date{}

\begin{abstract}
In 1861, Becquerel analyzed the time-resolved luminescence and formulated an empirical hyperbolic-type decay function, which was later named Becquerel decay law. Since then, studies about hyperbolic decays of luminescence have been carried on in different physical contexts. In this paper we generalize the Becquerel decay law by using a Mittag-Leffler function with a logarithmic argument, and discuss its physical interpretation in light of recently published results.

\end{abstract}

\maketitle

\section{Introduction}\label{sec1} 
The French physicist Edmond Becquerel not only designed his phosphoroscope to realize the first pioneering time-resolved photoluminescent experiments, but also found an empirical function for improving the fit previously obtained with one or a sum of two exponential functions. Starting from the phosphorescent intensity measured in inorganic solid samples, Becquerel modeled the following normalized empirical decay
\begin{equation*}
    I(t)=\frac{1}{\Big(1+\frac{t}{\tau_0}\Big)^2},
\end{equation*}
which was later generalized in
    \begin{equation}\label{bec}
    I(t) = \frac{1}{\Big(1+\frac{t}{\tau_0}\Big)^p},
    \end{equation} 
after placing other samples between the two rotating disks of his phosporoscope as the alkaline-earth sulfides. In \eqref{bec}  the intensity is normalized ($I(0)=1$), the Becquerel parameter varies between $1\leq p\leq 2$ and $\tau$ is a real parameter with the dimension of time. This equation is also known as \emph{Becquerel decay function}  \cite{ber_ber}, and includes both the hyperbolic decay ($p = 1$) and the \textit{squeezed hyperbola} ($1<p\leq2$), i.e. a function decaying faster than an hyperbola. More recently, Alvermann et al. \cite{yogi} showed how many physical systems particularly relevant to biology and bio-medicine seems to follow hyperbolic more than exponential decay laws. Anomalous behaviour in photoluminescence experiments can be observed only in the very long rung, as  the photoluminescent emission is quite satisfactory described by a bi-exponential model that is valid in the short run (hundreds of hours). In particular, it should be also noted that sample thermal treatment such as the annealing influences positively the intensity of the photoluminescent emission, flattening eventual abrupt dampings and unexpected plateau in photoluminescence, that becomes really difficult to observe \cite{c1,c2,c3,c4,lattanzi}. 

As can be seen from the genesis of the Becquerel decay function, so far the research has been lead on the basis of the experimental measurements and results. In this context, we suggest reversing course and trying to anticipate the experiments and their consequent modelling requirements by filling a theoretical gap. In fact, while the hyperbolic decay law and the compressed (or squeezed) hyperbola have already been taken into consideration, a \textit{stretched hyperbolic decay law}  has not yet been defined. Our approach, based on the analysis of the fractional differential equation, generalizes the classical decay equation with a time-dependent rate. Here we embed the time-variable rate by using the Caputo time-fractional derivative of a function with respect to another one (see \cite{jap} and \cite{almeida} for further details). From now on, we call \textit{stretched hyperbola} or \textit{stretched hyperbolic decay} the normalized phosphorescent emission in terms of the Mittag-Leffler function
\begin{equation}
    E_{\nu,\beta}(-\lambda t) =\sum_{k=0}^\infty \frac{(-\lambda t)^{\nu k}}{\Gamma(\nu k+\beta)},
    \end{equation}
that is proved to be useful in a large number of applications spanning different fields of the applied sciences, as illustrated in the recent monograph \cite{main}. More specifically, the Mittag-Leffler was introduced by Berberan-Santos et al. \cite{ber_ber,ber} to analyze luminescence decay with underlying distribution, and Lemes et al. \cite{lemes} to describe non-exponential chemical effects. Moreover, an application of Mittag-Leffler functions in the field of radiative transfer was recently suggested in \cite{Casa,Casa2}.
 It will be shown that the approach we propose not only restores a sort of \textit{symmetry} with the Kohlrausch-Williams-Watts (KWW) function, another well-known function used in photoluminescence \cite{c1,c2,lattanzi,ber1, c10,lukichev}, but also generalizes the Becquerel decay law in order to improve the agreement with experimental data when anomalous behaviours emerge.

The paper is organized as follows. In Section \ref{sec2}, we summarize the differential equations from which the decay law \eqref{bec} is derived, defining the context for the introduction of the fractional integro-differential equation that allows to bridge the hyperbolic law \eqref{bec} with its stretched version for the phosphorescence decay. The main result of the paper is presented in Section \ref{sec3}, where the stretched hyperbola in terms of the Mittag-Leffler function is introduced. The function and its behaviour is described in Sect. \ref{sec_ml_function}, whereas in Section \ref{sec4} we discuss our \textit{generalized (stretched) Becquerel decay} law from a physical point of view. Finally, concluding comments and remarks are summarized in  Section \ref{sec5}.

\section{The Becquerel function}\label{sec2}
The intensity degradation decay function $I(t)$ is proportional to the decay of the number of the emitting centers according to the following relation:
    \begin{equation}\label{ratio}
    I(t):=\frac{N(t)}{N_0},
    \end{equation} 
where $N(t)$ denotes the number of emitting centres at a time $t$ and $N_0=N(0)$ is their initial number. In other words, this ratio \eqref{ratio} defines the relaxation function of the luminescent system considered and the time-resolved photoluminescence.
The decay law \eqref{bec} can be derived mathematically by using two different approaches in modelling the time-evolution decay of the number of luminescent centers $N(t)$. Both these approaches are rooted in the physico-chemical reactions, allowing to deepen the physical meaning of the photoluminescent process.

As illustrated in \cite{ber1}, the first approach is based on a non linear differential equation
\begin{equation}\label{double_reactants}
        \frac{dN}{dt}= -k N^{1+\frac{1}{p}},
        \end{equation}
    which is reduced, when $p=1$, to
        \begin{equation}\label{double_reactants_1}
        \frac{dN}{dt}= -k N^2.
        \end{equation}
    According to the notation introduced in Section \ref{sec1} $k$ is equal to $\frac{p}{\tau_0}$, leading to
        \begin{equation}\label{double_reactants_2}
        \frac{dN}{dt}= -\frac{p}{\tau_0} N^2.
        \end{equation}\\
    Equation \eqref{double_reactants_2} describes a decay process where two reactants (electrons and holes) are involved in equal concentration, being a special case of the more general equation 
    \begin{equation}
    \frac{dN_1}{dt}=-kN_1N_2,
    \end{equation}
where $N_1$ and $N_2$ denote the concentrations of two different species. It's worth to note that the parallelism with such a physico-chemical approach turned out to be very useful in the interpretation of experimental data \cite{c1,lattanzi}.

The second approach used to retrieve the hyperbolic decay \eqref{bec} considers a first-order kinetic equation. Here, the reaction rate of decay is time-dependent and it is proportional to the number of emitting centers at that instant. The first-order kinetic equation has the form
        \begin{equation}\label{decay}
        \frac{dN}{dt}= -w(t) N,
        \end{equation}
    with a time-dependent rate
        \begin{equation}
        w(t)= \frac{p/\tau_0}{\Big(1+\frac{t}{\tau_0}\Big)},
        \end{equation}
    that can be reduced as follows when $p=1$:
        \begin{equation}
        w(t)= \frac{1/\tau_0}{\Big(\frac{t}{\tau_0}+1\Big)}.
        \end{equation}

\section{An integro-differential equation with logarithmic kernel for phosphorescence}\label{sec3}

In the recent paper \cite{jap}, an integro-differential operator with logarithmic kernel has been introduced in the context of correlated fractional negative binomial processes. This operator can be seen as a particular specialization in the more general definition of the Caputo fractional derivative of a function with respect to another function (see, e.g., \cite{almeida} and the references therein), and plays the role of a fractional evolution operator.

Using the same notation as in \cite{jap}, the time-evolution operator $\widehat{O}^t_\nu$ is defined as
    \begin{eqnarray}
    \nonumber &\widehat{O}^t_\nu f(t) =\displaystyle
    \frac{1}{\Gamma(n-\nu)}\int_{\frac{1-a}{b}}^{t}	\ln^{n-1-\nu}\left(\frac{a+bt}{a+b\tau}\right)\times\\
    \label{Def} &\displaystyle \times\bigg[\left(\Big(\frac{a}{b}+\tau\Big)\frac{d}{d\tau}\right)^n f(\tau)\bigg]\frac{b}{a+b\tau}d\tau,\nonumber
    \end{eqnarray}
for $n-1 < \nu < n$.
A relevant property of this operator is given by the following result (\cite{jap}, p.1057 for further details):
    \begin{equation}\label{pr}
    \widehat{O}^{t}_\nu \ln^\beta (a+bt) = \frac{\Gamma(\beta+1)}{\Gamma(\beta+1-\nu)}\ln^{\beta-\nu}(a+bt)
    \end{equation}
for $\nu \in (0,1)$ and $\beta>-1 \setminus{\{0\}}$. 
It can be proved by simple calculations that the composed Mittag-Leffler function 
    \begin{equation}\label{malt}
    E_{\nu,1}(-\ln^\nu(a+bt))= \sum_{k=0}^\infty\frac{(-\ln^\nu(a+bt))^k}{\Gamma(\nu k+1)},
    \end{equation}
    solves the following integro-differential equation 
    \begin{equation}\label{decay1}
     \widehat{O}^{t}_\nu f(t) = -f(t).
    \end{equation}
This equation can be considered as a generalization of the decay equation \eqref{decay}, where the time-dependence of the rate is involved in the memory kernel. By using this approach, both (logarithmic) memory effects and a time-dependent rate are taken into account. As a relevant remark, we observe that in the case $\nu = 1$, we have that 
\begin{equation}\label{operator_definition}
\widehat{O}^t_1 f(t) =\left(\Big(\frac{a}{b}+t\Big)\frac{d}{dt}\right) f(t)
\end{equation}
and thus, from the integro-differential equation \eqref{decay1}, we immediately recover the first-order kinetic equation
    \begin{equation}
    \frac{df}{dt}=-\frac{1}{\frac{a}{b}+t}f(t),
    \end{equation}
    that coincides with equation \eqref{decay} when $a=1$, $b=\frac{1}{\tau_0}$.
This result gives a first validation of the proposed approach, as the integro-differential equation \eqref{decay1} is introduced as a generalization of the well-known formula \eqref{decay}.\\

    \section{Becquerel decay generalization by means of Mittag-Leffler functions}\label{sec_ml_function}
     
      On the basis of the previous analysis and using $a=1$, $b=\frac{1}{\tau_0}$ in \eqref{operator_definition}, the solution to the integro-differential equation \eqref{decay1} gives the  following generalization of the Becquerel hyperbolic decay:
     \begin{equation}\label{n2}
      \frac{N(t)}{N_0} = E_{\nu,1}\Big(-\ln^\nu\Big(\frac{t}{\tau_0}+1\Big)\Big),
      \end{equation}
	where 
    \begin{equation}
    E_{\nu,1}\Big(-\ln^\nu\Big(\frac{t}{\tau_0}+1\Big)\Big)= \sum_{k=0}^{\infty} \frac{\left(-\ln^{\nu }\Big(\frac{t}{\tau_0}+1\Big)\right)^k}{\Gamma(\nu k +1)}
	\end{equation}
is the composition of the Mittag-Leffler function with the logarithmic function. The solution \eqref{n2} results to be normalized as required to facilitate the comparison between data collected under different environmental conditions. 
\begin{figure}
    \centering
    \includegraphics[scale=0.8]{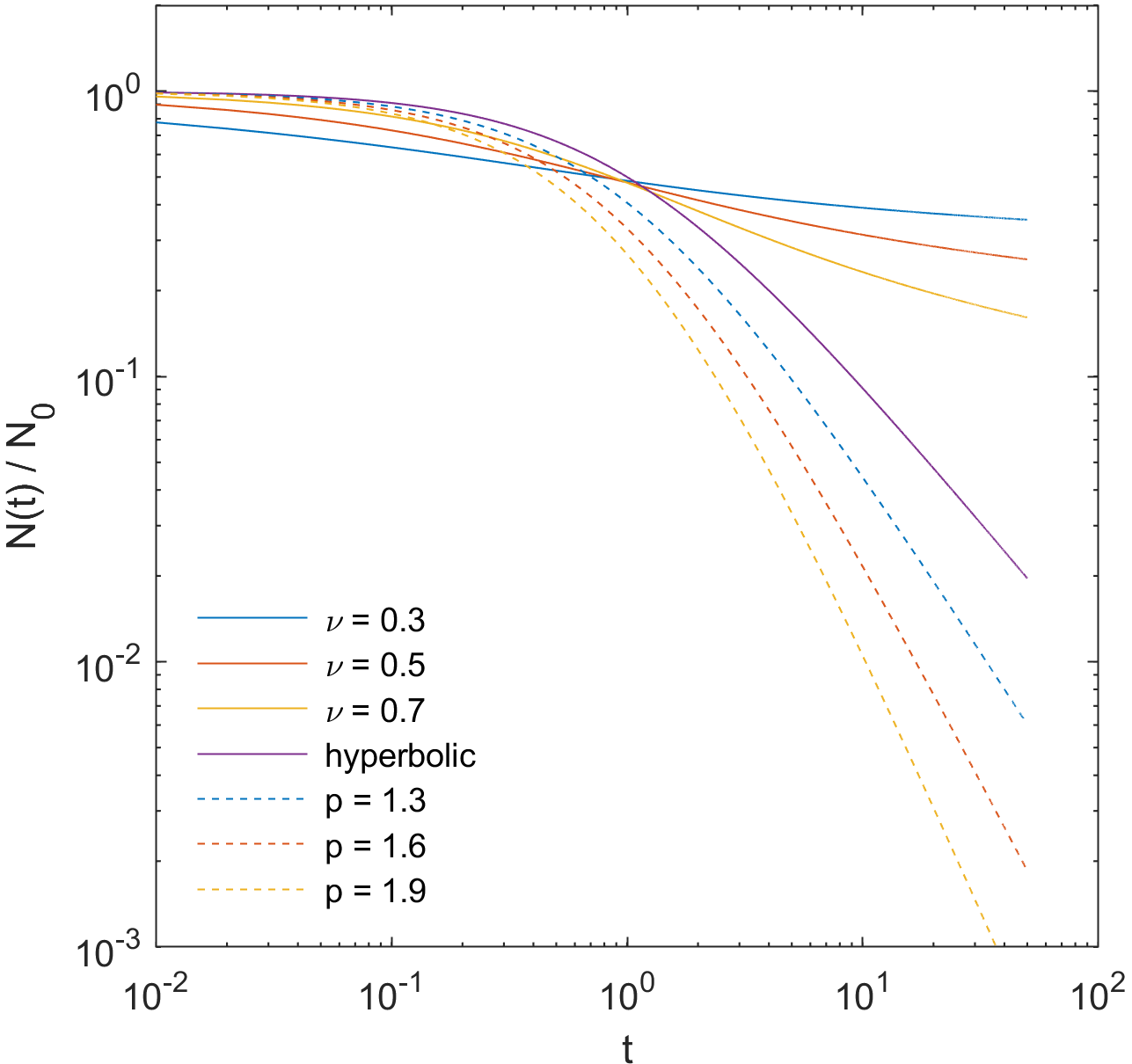}
    \caption{Log-log plot of Eq. \eqref{n2} (full lines) and \eqref{bec} (dashed lines) as a function of time $t$ for different value of $\nu$ and $p$, respectively.}
    \label{fig:01}
\end{figure}

Figure \ref{fig:01} reports in a single log-log plot the time-resolved normalized photoluminescent emission as a function of time $t$ for different value of $\nu$ and $p$. The solid purple line represents the hyperbolic Becquerel decay \eqref{bec} with $p=1$, the dashed lines are the squeezed hyperbolas \eqref{bec} defined for $1<p\leq2$ and the solid lines show the behaviour of the stretched hyperbola in terms of Mittag-Leffler function with logarithmic argument \eqref{n2}. Here, in this last case $\nu$ ranges in $(0,1)$. As expected, while the \textit{squeezed hyperbola} \eqref{bec} describes phenomena decaying faster than an hyperbola, the \textit{stretched} version \eqref{n2} we presented here applies to the opposite case. As observed in Section \ref{sec3} for the integro-differential equation \eqref{decay1}, the hyperbolic type decay is recovered when $\nu$ is set equal to $1$:
    \begin{equation}
    \frac{N(t)}{N_0}= \frac{1}{e^{-\ln\Big(\frac{t}{\tau_0}+1\Big)}}= \frac{1}{\frac{t}{\tau_0}+1}
    \end{equation}

In principle, it is possible to generalize the Bacquerel decay low by choosing among a large number of special and empiric functions. Anyway, Figure \ref{fig:01} clearly shows that the Mittag-Leffler function is able to describe the power-law-type decay expected when dealing with anomalous (i.e. non-Debye) relaxation and extinction processes (see, e.g. \cite{main}, \cite{maione}), and can be considered as a natural choice to describe the normalized photoluminescent decay for a stretched time-resolved luminescence (as also highlighted in \cite{ber}). It is worth to mention that unlike what reported in \cite{ber_ber}, we considered here a Mittag-Leffler function with a logarithmic argument because it allows for a better physical interpretation.

\section{Physical interpretation of the result}\label{sec4}

In the previous Section the Becquerel decay law was extended to describe a stretched hyperbolic behaviour, but its physical interpretation still needs to be clarified. A strategic approach to frame the physical meaning of photoluminescent empirical functions has been designed in \cite{lattanzi}, where it was applied to the Kohlrausch-Williams-Watts function. Despite the different function used to model the fluorescence, the methodology and the results are quite general and can be applied to other cases.

Firstly, we need to \textit{zoom} on the solution describing the time-resolved luminescence through a series expansion. Since the photoluminescence is modeled in the short run as a bi-exponential model, we can just consider the zero-th and the first order:
\begin{equation}\label{approx_solution}
\frac{N(t)}{N_0}=1-\frac{\ln^{\nu }\Big(\frac{t}{\tau_0}+1\Big)}{\Gamma(\nu +1)}.
\end{equation}
For the hyperbolic case $\nu=1$, our approximated solution \eqref{approx_solution} is identical to the time-dependent, non-exponential decay current observed in superconductors when the induced and resistive voltages are balanced (see, e.g., \cite{solid_art} and \cite{solid_book} p.113). In this context, the expression for the current can be obtained only after making a number of approximations in order to integrate exactly the equation governing the decay current:
\begin{equation}
\begin{split}\label{current_eq}
    -L\frac{di(t)}{dt}&=i(t)R(t)\\
    &={i(t)}R_ne^{\sqrt{(i(t)/i_0-1)^3}},
\end{split}
\end{equation}
that coincides with the differential equation \eqref{decay} if the rate $w(t)$ is equal to
\begin{equation}\label{w_coeff}
w(t)=\frac{R_ne^{\sqrt{(i(t)/i_0-1)^3}}}{L}.
\end{equation}
Here $i(t)$ and $i_0$ are the current at time $t$ and the reference current, respectively, while $L$ is the inductance and $R_n$ the normal resistance. In order to obtain an analytical solution of \eqref{current_eq} we need to perform a couple of approximations, reducing  the current dependence in the exponent of \eqref{w_coeff} from $\frac{3}{2}$ to 1 and considering the prefactor in the right hand side of \eqref{current_eq} as a constant:
\begin{equation}\label{solution_nu_1}
\begin{split}
    -L\frac{di(t)}{dt}&=i(t)R(t)\\
    &={i(t)}R_ne^{{-(1-i(t)/i_0)}},
\end{split}
\end{equation}
whose solution is
\begin{equation}\label{current_solution}
\frac{i(t)}{i_0}=1-{\ln^{ }\Big(\frac{t}{\tau_0}+1\Big)}.
\end{equation}
The current decay \eqref{solution_nu_1} corresponds to our approximated solution \eqref{approx_solution} for $\nu=1$.
At the beginning, when $i(t=0)=i_0$, the time-dependent resistance $R(t)$ is equal to $R_n$ and cause a rapid decay of current; as the current starts to decrease, the function in the exponent is no longer zero and the resistance decreases, leading to a slower current decay that follows a (non-exponential) logarithmic behaviour.

It is worth to highlight that while solution \eqref{current_solution} is retrieved by a number of phenomenological approximations of the decay rate \eqref{w_coeff} in the description of the induced current \cite{solid_art,solid_book}, our solution \eqref{n2} and its series approximation \eqref{approx_solution} naturally emerge by  generalizing the hyperbolic decay law to be valid in the stretched range. In other words, our integro-differential equation offers a way to solve the physical problem of the induced current when it exceeds the critical current density on the surface without making any approximation \cite{das}.

The mathematical and formal analogy between induced current and photoluminescent emission is also to be found in the physical meaning underlying the process dynamics. The exponent in \eqref{w_coeff} describes a current-dependent barrier height that is almost zero at the beginning and causes a slower current decay when starts to increase (i.e., the resistance decreases, as reported in \cite{solid_art,solid_book}). In photoluminescence processes, the role of {this increasing} barrier can be understood in the light of the reduced (or effective) mass, that can be related to the inertia of the system preventing the photoluminescent emission \cite{lattanzi}. {The reduced mass can be associated to the presence of the cloud of virtual particles around the emitting centers, that deform the surrounding environment creating a distribution of relaxation times (that can be modeled as traps) and leading to a non-hyperbolic, anomalous behaviour.} In particular, it was found that there are two different behaviours associated to the reduced mass depending on whether the function is complete monotone or simple monotone.\\ According to Theorem 2 in \cite{miller}, the photoluminescence described by \eqref{n2} is a complete monotone function, because the Mittag-Leffler function is a complete monotone function for $\nu\in(0,1)$ as proven in \cite{pollard,monotonic} and the logarithm is a non-negative function with a complete monotone derivative. {As shown in \cite{lattanzi}}, it implies that the corresponding reduced mass is increasing and therefore, the barrier is also increasing, confirming the result obtained {in \cite{solid_art,solid_book}.}     
{The increasing barrier implies that the dynamics is ruled by an exclusive damping mechanism. Assuming that the system experiences such type of single dynamics, the function \eqref{n2} effectively describes a phosphorescence emission as explained here below.\\}
After an initial photoexcitation, the atoms responsible for the luminescence are activated: the incident radiation, usually in the ultraviolet range, creates excitons (bound states composed by an electron and an hole) that play the role of the emitting centers; the recombination of these charges produces the observed luminescence.
The photoexcited electron initially populates the singlet state and then it can decay emitting or not radiation to the ground state. This radiation is the so-called photoluminescence. After the rapid initial decay of photoluminescence, other decay pathways are activated. A fraction of photoexcited electron follows another decay path called intersystem crossing and starts to populate the triplet state. According to this framework, the triplet state is a sort of trap-state where the excited electrons cannot decay immediately via phosphorescence due to the small value of the decay rate $K_\mathcal{T}=\frac{1}{\tau_T}$.
Now, the thermally activated processes take place favouring the emission as a thermally activated delayed fluorescence (TADF) and room-temperature phosphorescence. However, considering that the function is completely monotone, only decay pathway should be considered as the non-radiative annihilation via recombination with more active species as molecular oxygen $O_2$, or the phosphorescent emission. And it is perfectly framed in the context in which we derived the normalized photoluminescent decay function \eqref{n2}: it describes the phosphorescence.\\ The de-trapping is generally described by an Arrhenius law of this form
\begin{equation}\label{risc_rate}
K_{\mathcal{T}}\simeq e^{-\frac{\Delta E}{k_B T}},
\end{equation}
where $\Delta E$ is the energy gap between the metastable and the excited triplet state, $k_B$ is the Boltzmann constant, and $T$ is the temperature of the system which is time-dependent. This decay rate plays the role of a de-trapping rate, i.e. the rate at which the electrons are recombined and the phosphorescence arise.

\section{Conclusion and final remarks}\label{sec5}
The Becquerel function was firstly introduced in the XIX century to describe the phosphorescence emitted by inorganic solid samples, and includes both the hyperbolic and the squeezed hyperbolic decays. In this paper, we generalized the Becquerel decay law via a Mittag-Leffler function with logarithmic argument to include also stretched decays, making an analogy with the well-known Kohlrausch-Williams-Watts functions. This result was further analyzed considering its series expansion stopped at the first order, that allowed to make a comparison with recently published results about the description of photoluminescence in complex systems.

\section*{Acknowledgments}
A.L. was supported by the NCN research project OPUS 12 no. UMO-2016/23/B\\/ST3/01714 and by the National Agency for Academic Exchange NAWA project: Program im. Iwanowskiej PPN/IWA/2018/1/00098.

\section*{CRediT authorship contribution statement}
All the authors contribute equally to the work.

\end{document}